# Will 3552 Don Quixote Escape from the Solar System?

**Suryadi Siregar**[1,2]

[1]Bosscha Observatory, Lembang 40391, Indonesia
[2]Astronomy Research Division, Faculty of Mathematics and Natural Sciences,
Institut Teknologi Bandung, Bandung 40132, Indonesia

**Abstract.** Asteroid 1983 SA, well known as 3552 Don Quixote, is one of Near Earth Asteroids (NEAs) which is the most probable candidate for the cometary origin, or otherwise as Jupiter-Family-Comets (JFCs). The aim of this study is to investigate the possibility of 3552 Don Quixote to be ejected from the Solar System. This paper presents an orbital evolution of 100 hypothetical asteroids generated by cloning 3552 Don Quixote. Investigation of its orbital evolution is conducted by using the SWIFT subroutine package, where the gravitational perturbations of eight major planets in the Solar System are considered. Over very short time scales (~220 kyr) relative to the Solar System life time (~10 Gyr), the asteroid 3552 Don Quixote gave an example of chaotic motion that can cause asteroid to move outward and may be followed by escaping from the Solar System. Probability of ejection within the 220 kyr time scale is 50%.

## 1 Introduction

Since 1984 asteroid study has been conducted at the Bosscha Observatory. Presently, there are two main streams in our research, i.e. dynamical and physical studies. The aim of this study is a continuation of previous long-term study of NEAs. Two main theories are introduced to explain the origin of the asteroids (Wetherill [1]). An earlier theory assumes the existence of a parent body, between Mars and Jupiter, which was broken to pieces due to perturbation by Jupiter. Some of the pieces were ejected to libration point L4 and L5, called the Trojan asteroids, and others became the MBAs (Main Belt Asteroids). Nowadays the largely accepted theory considers asteroids as small condensations of primitive solar nebula that were unable to form into a single body, due to the presence of Jupiter. Jupiter and Saturn perturbed strongly their semimajor axis and its influence decreases when the planets' heliocentric distances increased to allow for protoplanet migration [2]. In addition, Leinhardt & Richardson [3] developed a self-consistent planetesimal collision model that includes fragmentation and accretion of debris. The numbers and masses of protoplanets and the time required to grow a protoplanet depend strongly on the initial conditions of the disk. The elasticity of the collision, does not significantly affect planetesimal growth over longtime scale. Most of the asteroids move between Mars and Jupiter and collisions occur frequently. These collisional destructions occurred so often during the lifetime of the Solar System, that practically all the asteroids we now see are fragments of their original parent bodies. Some may be found in unstable zone like those of the Kirkwood gaps, in which they became the sources of Apollo-Amor-Aten asteroids (AAAs). This group is the main reference in the classification of NEAs. Study on the elliptical motion and perihelion distribution yields the perihelion that occupies the whole region of sky. The perihelion distance q spreads in the range 0.18 AU < $q$ < 1.38 AU with the strong concentration from 0.63 up to 1.38 AU. The average ⟨ $q \sin i$ ⟩ is 0.24 AU where $i$, is the inclination of orbit (Siregar [4,5]). At the time being Atira asteroids are considered as new members of NEAs. The definitions, orbital criteria and briefs description of NEAs group such as Atiras, Atens, Apollos, and Amors are briefly discussed in Siregar [6].

Asteroid 3552 Don Quixote was discovered by Paul Wild, on September 26, 1983, and has an alternate name 1983 SA. According to its orbital elements, shown in Table 1, and orbital criteria, this asteroid is known as Mars-crosser asteroid, Amor IV asteroid and Jupiter-crosser asteroid. 3552 Don Quixote is the most probable candidate for the cometary origin (Lupishko, et al. [7]). On the other hand, in certain case some NEAs are dynamically present as comets, for example: 1997 SE5, 1982 YA, 2002 RN38, and 2002 VY94 (Siregar [8]). At the same time many of the cometary candidates have physical properties that are inconsistent with our current understanding of cometary nuclei such as albedos and (or) unusual spectra. (For reviews on the relation of AAA asteroids and comets, see Siregar [9] and Fernandez, et al. [10]). All these properties clearly indicate that the MBA is the principal source of NEAs, and comet nuclei contribution to the total NEAs population does not exceed 10% (Yoshida & Nakamura [11]).

To obtain a comprehensive view of orbital evolution 3552 Don Quixote, this work undertakes a numerical integration of the Solar System motion for over 250 kyr. In this time interval, we concentrate only on gravitational perturbation of planets, because the Sun is still in the main-sequence of Herztprung-Russel diagram such that its radiation does not differ from the present time. The influences of Sun's radiation with diameter 19 km do not play an important role. The Solar System gravitation is strongly dominant compared to thermal effects (Yarkovsky forces and YORP torques, see Bottke, et al. [12] for review). Meanwhile, in general rule, the non-gravitational forces may have important effects at small perihelion distances ($q \leq 1$ AU), and the non-gravitational forces have little influences for $q > 1.5$ AU, and so can be ignored (Emel'yanenko & Bailey [13]).

## 2  Orbital elements

By using the observational data in period 1983 up to 2009 the orbital elements of 3552 Don Quixote has been already catalogued. In this paper all the information of the orbital elements are taken from http://ssd.jplnasa.gov.

**Table 1**  Orbital elements of 3552 Don Quixote as test particle. All of the elements refer to epoch JD 24455200.5 or 2010-Jan 04.0; see also Siregar [6]

| No | Element | Symbol | Value | Uncertainty ($1\sigma$) | Unit |
|----|---------|--------|-------|-------------------------|------|
| 1  | Eccentricity | $E$ | 0.7138803024084498 | $6.5373 \times 10^{-8}$ | |
| 2  | Semimajor axis | $A$ | 4.224923970817345 | $1.4119 \times 10^{-8}$ | AU |
| 3  | Perihelion distance | $Q$ | 1.20883396887755 | $2.7877 \times 10^{-7}$ | AU |
| 4  | Inclination | $i$ | 30.969330692119038 | $1.6838 \times 10^{-5}$ | Deg |
| 5  | Ascending node | $\Omega$ | 350.2673100000013 | $2.3096 \times 10^{-5}$ | Deg |
| 6  | Perihelion argument | $\omega$ | 317.1044852074207 | $2.4267 \times 10^{-5}$ | Deg |
| 7  | Mean anomaly at epoch | $M$ | 13.32372560374069 | $4.0456 \times 10^{-6}$ | Deg |
| 8  | Perihelion passage | $To$ | 2455083.1049045711 | $3.5394 \times 10^{-5}$ | JED |
| 9  | Sideral's period | $P$ | 8.68 | $4.353 \times 10^{-8}$ | Yr |
| 10 | Daily motion | $N$ | 0.113494738047304 | $5.689 \times 10^{-10}$ | deg day$^{-1}$ |
| 11 | Aphelion distance | $Q$ | 7.24101397275714 | $2.4199 \times 10^{-8}$ | AU |

To simulate the evolution of 3552 Don Quixote's trajectory we clone the orbital elements by generating random number from 0 to 1 according to each orbital elements with uncertainties shown in Table 1. By using these values, 100 clones are generated corresponding to hypothetical Don Quixote-like asteroids that follow Gaussian distribution. All orbital elements are cloned by using probability density function (*pdf*) according to Gauss formula:

$$pdf = A e^{-\left(\frac{x-\mu}{2\sigma}\right)^2} \tag{1}$$

where $\sigma$ is the uncertainty value of the orbital elements, $x$ is the value of the orbital elements, and $\mu$ is the mean of orbital elements or the 3552 Don Quixote's orbit itself. Normalization factor $A$ is taken equal to 1. Figure 1 presents the *pdf* of orbital elements of 100 clones asteroid Don Quixote-like. All of the distribution functions demonstrate the Gaussian form of uncertainty in the orbital elements.

## 3  Orbital Evolution Scheme

According to the N-body problem the acceleration vector of asteroid motions consists of three factors. The first one is the acceleration force due to the Sun, the second and the third ones are acceleration forces by planets to asteroid and planet to the Sun, respectively [14]. Dynamical evolution of the asteroid 3552 Don Quixote was determined by inspecting positions and velocities of the asteroid and all planets. This was done by employing numerical integration of general Cartesian N-body Newtonian equation of motion along the time-evolution. The asteroid is assumed to be massless. By taking a time step of 1/1000 yr ~ 1/240 of Mercury's period, and considering the perturbation of all planets in the Solar System, we deduced that relative accuracies of the integration attain maximum at an order of 10-6 in total energy and angular momentum which are appropriate enough for a conservative system. The SWIFT package used in this work is designed to integrate a set of mutually gravitationally interacting bodies together with a group of test particle under the gravitational influence of the massive bodies but does not affect each other. In this work Regularized Mixed Variable Symplectic (RMVS) method is used. This also considers close approaches between test particles and planets [15].

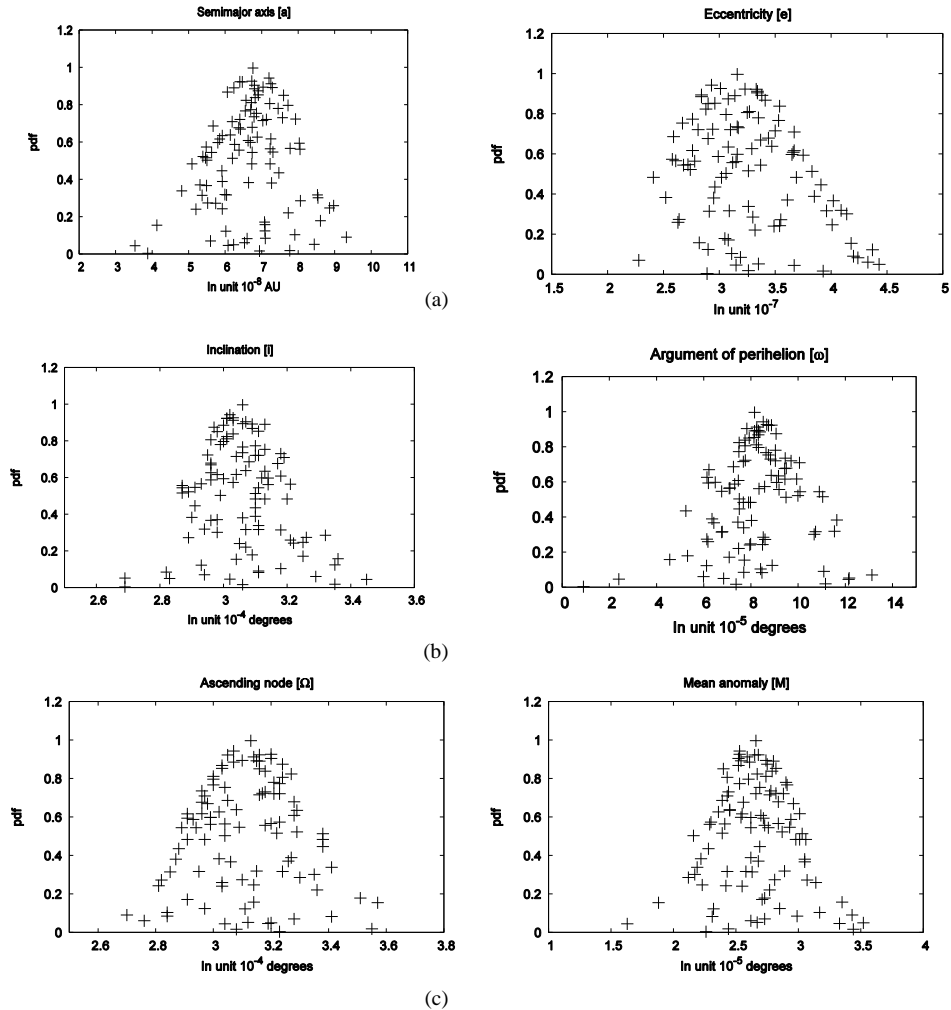

**Figure 1** Probability density function of 100 clones of 3552 Don Quixote. **(a)** Left: The semimajor axis, the origin of $x$-axis in the graph is 4.2249239 AU. Right: The eccentricity, the origin of $x$-axis in the graph is 0.713880. **(b)** Left: The inclination, the origin is 30.969 degree. Right: The argument of perihelion, the origin is 317.1044 degree. **(c)** Left: The ascending node, the origin is 350.267 degree. Right: The mean anomaly, the origin is 13.3237 degree.

For initial conditions that are in agreement with our present knowledge of the parameters of the Solar System, we used the orbital data of principal planets in our Solar System issued by NASA (http://ssd.jplnasa.gov). The compilation of the orbital elements and masses of eight planets as perturbing bodies at epoch JD 24455200.5 (2010-Jan 04.0) also can be seen in Siregar [6].

## 4     Results and Discussion

The summary of our computational results for eccentricity, semimajor axis, perihelion distance, inclination, ascending node, argument of perihelion, period in year, daily motion, and aphelion distance are presented in Table 2.

**Table 2**   Orbital elements of 3552 Don Quixote now and 220 kyr later.

| No | Element | Now | Later |
|---|---|---|---|
| 1 | $e$ | 0.7139 | 0.9741 |
| 2 | $a$ (AU) | 4.2249 | 142.2773 |
| 3 | $q$ (deg) | 1.2088 | 3.6835 |
| 4 | $i$ (deg) | 30.9693 | 24.7523 |
| 5 | $\Omega$ (deg) | 350.2673 | 71.2154 |
| 6 | $\omega$ (deg) | 317.1045 | 304.7182 |
| 7 | $P$ (year) | 8.6800 | 1697.0850 |
| 8 | $n$ (deg day$^{-1}$) | 0.1135 | 0.0006 |
| 9 | $Q$ (AU) | 7.2410 | 280.8712 |

Over 220 kyr the mass of the Sun is relatively constant, and also the orbit of planets both outer and inner are essentially not affected by the Solar System gravitation. The constancy of semimajor axes of planets were demonstrated in previous studies (see, for example, Siregar [6]).

These situations are completely different for 3552 Don Quixote, whose orbital eccentricity and semimajor axis changes rapidly. We found that the orbit is not stable and chaotic as can be seen in Figure 2.

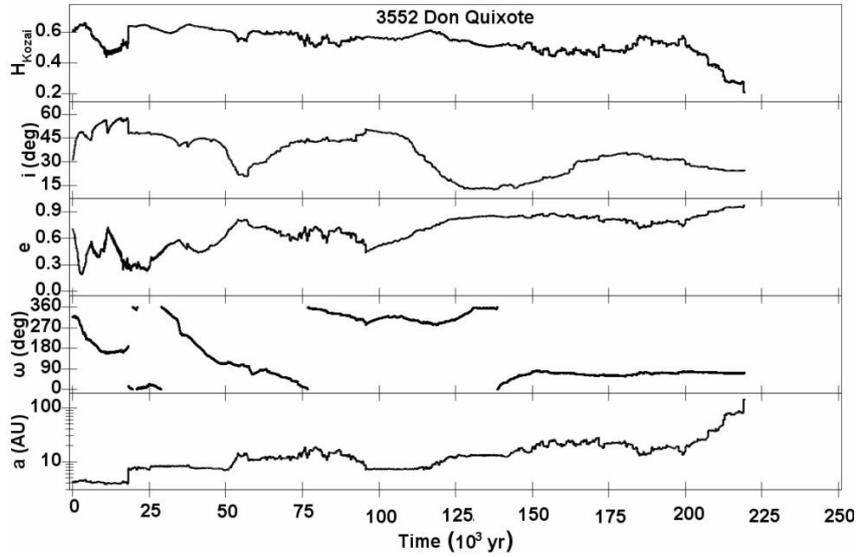

**Figure 2** Evolutions of semimajor axis, *a* (in AU), argument of perihelion in degree, eccentricity (*e*), inclination (*i*) in degree, and the Kozai resonance, $H_{Kozai}$. Integration time-step is $10^{-3}$ year by including all of planet's perturbations. A transfer of orbital shape from elliptic to parabolic exist at time interval $t \sim 220$ kyr (exact number is $t = 219301$ year), where the semimajor axis and eccentricity are 142.2 AU, and 0.9741, respectively. The integration of orbital elements is stopped at $t = 219301$ year.

Jupiter is the planet with the largest mass in the Solar System. Therefore, it is quite reasonable when we consider the dominant perturbation on 3552 Don Quixote caused by Jupiter. According to restricted three-body problem, to distinguish asteroids from Jupiter family comets, we calculate the Tisserand's invariant, $T_j$ by using the expression:

$$T_J = \frac{1}{a} + 2\sqrt{a(1-e^2)}\cos i \qquad (2)$$

where semimajor axis *a* is in unit of Jupiter's semimajor axis. Asteroids and comets have $T_j > 3$ and $T_j < 3$, respectively. This general rule is not strictly fulfilled in all cases. Tisserand's invariant of the system Sun-Jupiter-Don Quixote is $T_j = 2.184$. This implies that gravitational perturbations from other planets in the Solar System do not affect at all. The 3552 Don Quixote is still in Jupiter Family Comets (JFCs). Physically it is an asteroid but dynamically it is a comet. This phenomenon is also shown by 2P/Encke with $T_j = 3.018$, and 107P/Wilson-Harrington with $T_j = 3.086$, which both physically are comets (extinct comets) but dynamically are asteroids.

The second term in the expression (2) is known as the Kozai resonance (Kozai, [16]),

$$H_{Kozai} = \sqrt{a(1-e^2)}\cos i \quad (3)$$

For comparison, see also the Kozai resonance for NEAs with semimajor axis smaller than 2 AU (Michel & Thomas [17]). The profile of asteroid track is presented in Figure 3.

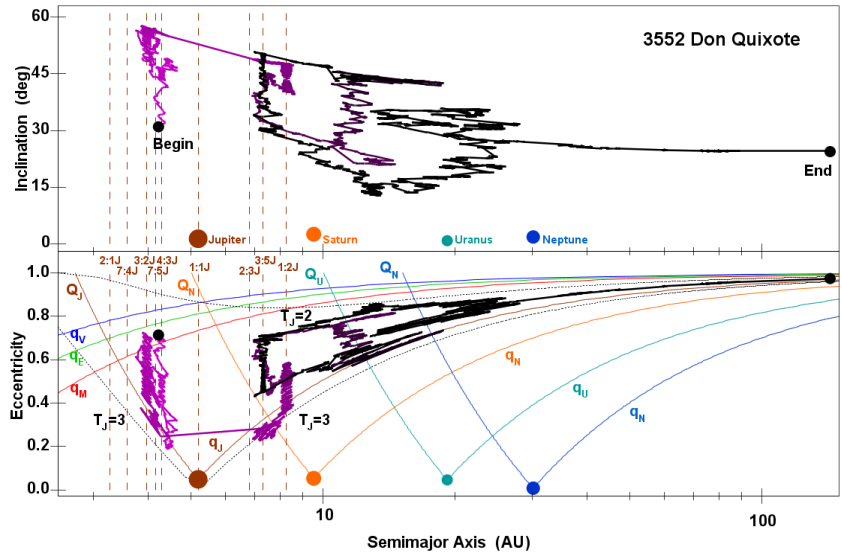

**Figure 3** Orbital track of asteroid 3552 Don Quixote over 220 kyr, gradual color changes from grey to black denotes time-evolution. $T_x$, $Q_x$, and $q_x$ represent Tisserand invariant, aphelion distance, and perihelion distance of perturber planet, respectively. Vertical broken-lines represent Mean-Motion-Resonance (MMR) by planet.

As displayed in Figure 3, during the orbital evolution of this asteroid over 220 kyr, we can see that this asteroid will be trapped in Mean-Motion-Resonance (MMR), that is temporarily trapped in MMR 4:3J, 3:2J, 7:5J and temporarily trapped in MMR 3:5J and it interacts at a boundary of Tisserand invariant Tj = 3. Furthermore, it will be temporarily trapped in MMR 1:2J, 3:5J and perturbed at Uranus aphelion, and again temporarily trapped in MMR 3:5J. In MMR mechanism the perturbation of other planets on semimajor axis of 3552 Don Quixote is not significant, if the semimajor axis of asteroid is constant, and the eccentricity *e*, and the inclination *i*, undergo oscillations such that $\sqrt{(1-e^2)}\cos i$ = constant. Therefore, if eccentricity *e* increases then the inclination *i* decreases, and vice versa. We do not find 3552 Don Quixote locked in the Kozai resonance. At the end of integration $H_{\text{Kozai}}$ decreases gradually (see Figure 2). This condition is also shown by the evolution of argument of perihelion where $\omega$, does not librate around $90^{\circ}$ or $270^{\circ}$. Finally the asteroid moves toward outside the Solar System by passing the area of Tisserand invariant in the range, $2 < T_j < 3$.

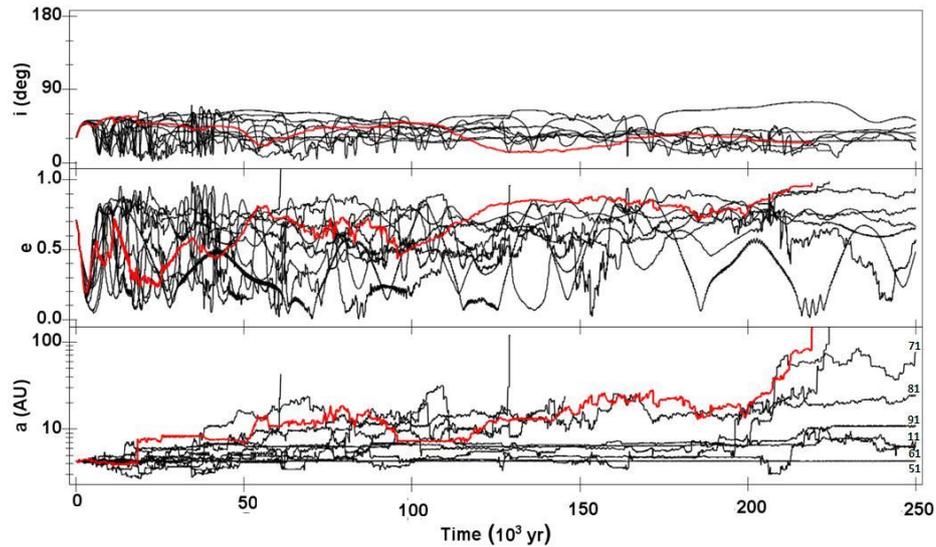

**Figure 4** Orbital evolutions of 10 hypothetical asteroids generated by cloning 3552 Don Quixote. The clone numbers 1, 21, 31, and 41 escapes from the Solar System, but numbers 11, 51, 61, 71, 81, and 91 are still in the Solar System. The red one is 3552 Don Quixote.

To fully illustrate the future status of 3552 Don Quixote, we demonstrate in Figures 4 to 6 the computational results of the orbital evolution of hypothetical asteroids, generated by cloning 3552 Don Quixote, for which the *pdf* of these asteroids are shown in Figure 1.

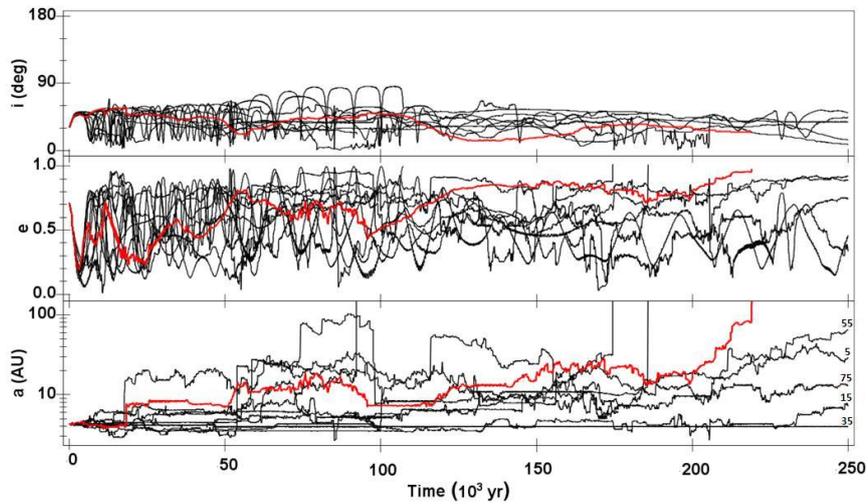

**Figure 5** Orbital evolutions of 10 hypothetical asteroids generated by cloning 3552 Don Quixote. The clone numbers 25, 45, 65, 85, and 95 escapes from Solar System, but numbers 5, 15, 35, 55, and 75 are still in the Solar System. The red one is 3552 Don Quixote.

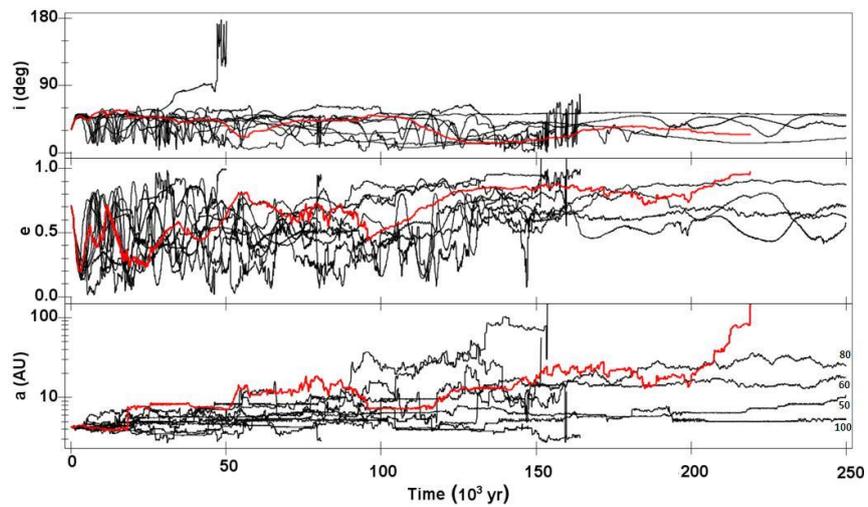

**Figure 6** Orbital evolutions of 10 hypothetical asteroids generated by cloning 3552 Don Quixote. The clone numbers 10, 20, 30, 40, 70 and 90 escapes from the Solar System, but numbers 50, 60, 80, and 100 are still in the Solar System. The red one is 3552 Don Quixote.

The computation on a relatively short time of a quantity, called Fast Lyapunov Indicators (FLI), allows to discriminate between ordered and weak chaotic motion and also under certain conditions between resonant and non resonant regular orbits (for detail explanations see Froeschle & Lega [18] and Lega & Froeschle [19]). Following Lega & Froeschle [19] FLIs of 3552 Don Quixote and some orbitals clones were examined. Figure 7 presents examples of variation of the FLI with time for 3552 Don Quixote and its clones.

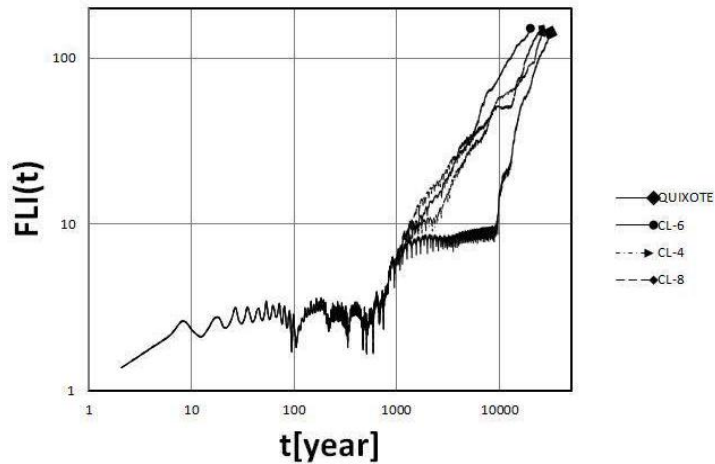

**Figure 7** FLI versus time (*t*) on log-log scale. 3552 Don Quixote demonstrates a dramatic change of FLI value from $t \sim 10$ kyr. Three other clones (CL- 4, CL-6 and CL- 8) show some characteristics in much earlier from $t \sim 800$ yr.

Figures 4 to 6 demonstrate that in a time interval of 250 kyr, the inclination, the eccentricity, and the semimajor axis of some hypothetical asteroids are unstable. Although the integration is done for a very short time compared to the Solar System time scale (~10 Gyr), it is found that almost 50% hypothetical asteroids are already ejected. The remaining clones are still in our Solar System; the majority of them show the tendency of the increasing eccentricities and the semimajor axes to increase. Otherwise the semimajor axes of asteroids numbers 35, 51, and 100 are relatively constant. If their eccentricities increase and inclination do not change significantly, then the perihelion distances should decrease, and it is possible for these asteroids to hit the Sun. Example of the sungrazing phenomenon for Jupiter-family is P/Encke, and for long period orbit (Halley-types), is P/Machholz which shows a collision with the Sun to occur within ~12 kyr (Farinella, et al [20], Bailley, et al [21]).

## 5 Conclusion

Asteroid 3552 Don Quixote is a clear example of the complexity of motion that can be exhibited by purely gravitating bodies in the Solar System. All planets have key roles to play in the evolution of 3552 Don Quixote. This asteroid also serves as an example of behavior chaotic that can cause asteroid to migrate outward, and may be followed by escaping from the Solar System.

As shown in Figures 4 to 6, we found that 51 among 100 clones remain in the Solar System for the time span adopted in this work. Approximately, it can be concluded that the chance of 3552 Don Quixote to escape from our Solar System is 50% over time interval of 220 kyr. We predict the future status of 3552 Don Quixote to be vanishing as a small body in the Solar System within 220 kyr.

### Acknowledgements

The author is indebted to Dr. Budi Dermawan for fruitful discussion and computer programming. A part of this work was presented at the Conference of the 2$^{nd}$ South-East Asia Astronomical Networking (SEAAN), Manila 16-19 February 2010. The travel grant from the Leids Kerkhoven-Bosscha Fonds is highly appreciated.